\documentclass[12pt,final]{article}
\usepackage{amsmath,amssymb,amsfonts,amsthm,slashed,braket}
\usepackage[all]{xy}
\usepackage{mathrsfs}
\usepackage{fancyhdr}
\usepackage{showkeys}

\newcommand\nc{\newcommand}
\nc\linesep{\bigskip}
\nc\newprob[1]{\marginnote{#1}[\parskip]}
\nc\bA{\mathbb A}
\nc\bC{\mathbb C}
\nc\bD{\mathbb D}
\nc\bR{\mathbb R}
\nc\bZ{\mathbb Z}
\nc\bQ{\mathbb Q}
\nc\bP{\mathbb P}
\nc\bV{\mathbb V}
\nc\bW{\mathbb W}
\nc\bG{\mathbb G}
\nc\mf\mathfrak
\nc\mc\mathcal
\nc\mb\mathbb
\nc\brac[1]{\langle#1\rangle}
\nc\abs[1]{\lvert#1\rvert}
\nc\norm[1]{\lVert#1\rVert}
\nc\onto{\twoheadrightarrow}
\nc\into{\hookrightarrow}
\nc\lto{\longrightarrow}
\nc\action{\curvearrowright}
\nc\on\operatorname
\nc\wbar\overline
\nc\what\widehat
\nc\wtilde\widetilde
\nc\nop\DeclareMathOperator
\nc\eps{\varepsilon}
\nc\tsym{\widetilde{\text{Sym}}}
\nc\oarrow[1]{\overset{#1}\to}
\nop\Hom{Hom}
\nop\End{End}
\nop\Aut{Aut}
\nop\im{Im}
\nop\id{id}
\nop\tr{Tr}
\nop\coker{coker}
\nop\Spec{Spec}
\nop\Jac{Jac}
\nop\Ext{Ext}
\nop\Tor{Tor}
\nc\op{\text{op}}
\nop\loc{Loc}
\nop\Frac{Frac}
\nc\ann{\text{ann}}
\nop\QCoh{QCoh}
\nop\Coh{Coh}
\nop\Sym{Sym}
\nop\Hilb{Hilb}
\nop\gr{Gr}
\nop\Tot{Tot}
\nop\Fl{Fl}
\nop\tGamma{\widetilde\Gamma}
\nop\tloc{\widetilde{\text{Loc}}}
\nop\rep{Rep}
\nop\proj{Proj}
\nc\oo[1]{\overset\circ{#1}}
\nop\ospec{\oo{Spec}}
\nop\oTot{\oo{Tot}}
\nop\Bl{Bl}
\nop\Comp{Comp}
\nop\Ho{Ho}
\nop\cone{Cone}
\nop\LKE{LKE}
\nop\RKE{RKE}
\nop\pd{pd}
\nop\cd{cd}
\nop\depth{depth}
\nop\ass{Ass}
\nop\supp{supp}
\nop\codim{codim}
\nop\holim{\underset{\lto}{holim}}
\nop\dlim{\underset{\lto}{lim}}

\nop\uHom{\underline{\Hom}}
\nop\Pic{Pic}
\nop\Cl{Cl}
\nop\Div{Div}
\nop\rank{rank}
\nop\Der{Der}
\nop\dimrel{dim.rel}
\nc\sHom{\mathscr Hom}
\nc\sExt{\mathscr Ext}
\nc\dto{\dashrightarrow}
\nop\rspec{\bf Spec}
\nop\Gal{Gal}
\nop\Ind{Ind}
\nop\Frob{Frob}
\nop\Fib{Fib}
\nop\ratdim{rat\ dim}
\nop\Mod{Mod}
\nop\rat{rat}
\nop\val{val}
\nop\Rep{Rep}
\nop\colim{colim}
\nop\ind{ind}

\theoremstyle{theorem}

\theoremstyle{remark}

\begin{document} 
%\maketitle

\centerline{\large{The Hodge-elliptic genus, spinning BPS states, and black holes}}
%\centerline{\large{Which genus has the largest p-ness?}}
\bigskip
\bigskip
\centerline{Shamit Kachru$^1$ and Arnav Tripathy$^2$}
\bigskip
\bigskip
\centerline{$^1$Stanford Institute for Theoretical Physics}
\centerline{Department of Physics, Stanford University, Palo Alto, CA 94305, USA}
\medskip
\centerline{$^2$Department of Mathematics, Harvard University}
\centerline{Cambridge, MA 02138, USA}
\bigskip
\bigskip
\begin{abstract}
We perform a refined count of BPS states in the compactification of M-theory on $K3 \times T^2$, keeping track of the information provided by both the $SU(2)_L$ and $SU(2)_R$ angular momenta in the $SO(4)$ little group. Mathematically, this four variable counting function may be expressed via the motivic Donaldson-Thomas counts of $K3 \times T^2$, simultaneously refining Katz, Klemm, and Pandharipande's motivic stable pairs counts on $K3$ and Oberdieck-Pandharipande's Gromov-Witten counts on $K3 \times T^2$.  This provides the first full answer for motivic curve counts of a compact Calabi-Yau threefold. Along the way, we develop a Hodge-elliptic genus for Calabi-Yau manifolds -- a new counting function for BPS states that interpolates between the Hodge polynomial and the elliptic genus of a Calabi-Yau. 

\end{abstract}

\tableofcontents

\section{Introduction and summary}

This paper is motivated by natural enumerative questions in physics and mathematics, related to degeneracies of rotating black holes on the one hand, and enumerative geometry of Calabi-Yau manifolds and allied questions in geometry and number theory on the other.  Therefore, we will provide a brief
introduction to both the physical and mathematical problems of interest.

\subsection{Entropy counts}

Understanding exact counts of BPS states in string vacua has been an ongoing project of significant interest for the past twenty years.
As one primary application, computing protected indices and increasing the string coupling until the states collapse into black holes is the predominant approach for obtaining microscopic accounts of black hole entropy.  The first work in this direction was that of Strominger and Vafa \cite{StromingerVafa} counting BPS states in type IIB string theory on $K3 \times S^1$ (or, equivalently, M-theory on $K3 \times T^2$). Since then, much effort has been devoted to refining these counts, often in the related four-dimensional string vacuum given by compactifying type IIA or IIB string theory on $K3 \times T^2$ (for reviews, see \cite{Senreview,Atishreview}); the counts of BPS states in the 4d and 5d theories are related via the 4d-5d lift, as explained in \cite{SSY}.
%the counts in the theories allowing for easy translation as they differ only by the center-of-mass degrees of freedom of the fully %wrapped D6 brane. 
In such 4d $\mc{N} = 4$ theories, we have two classes of BPS states -- those that preserve half of the supersymmetry, and those preserving only a quarter. The $1/2$-BPS states may be assumed to carry only electric charge in a suitable U-duality frame, whereas the $1/4$-BPS states are dyons, carrying both electric and magnetic charges.  As such, an unflavored count of $1/2$-BPS states should depend on only one quantum number, namely the single charge, while the count for $1/4$-BPS states takes slightly more care. The electric and magnetic charges $Q_e$ and $Q_m$ in the charge lattice transform under the action of the U-duality group, and there are precisely three U-duality invariant combinations, namely $Q_e \cdot Q_e, Q_e \cdot Q_m,$ and $Q_m \cdot Q_m$.  So, an accounting of the $1/4$-BPS states (without flavoring by any other information) should be a generating function in three variables.  A precise form for this function was first proposed by Dijkgraaf-Verlinde-Verlinde (DVV) in \cite{DVV}. We will review the formulae for the 1/2-BPS and 1/4-BPS counts in the next section.

While the existing story of 1/4-BPS counts therefore involves a partition function tracking three gradings, from at least two perspectives
it seems that a four-variable function would be more natural.  The first perspective is that of entropy counts.  The little group of a massive particle is $SO(4) \sim SU(2)_L \times SU(2)_R$.  In \cite{BMPV}, black hole solutions with two charges ($Q_F$ and $Q_H$) and two
angular momenta (in two two-planes in ${\mathbb R}^4$) are discussed.  In the limit that $K3$ is small compared to the circle, the degeneracy of supersymmetric ground states can be computed in an effective string theory (wrapping the $S^1$).  This theory
has $(4,4)$ supersymmetry, and the angular momenta map to the left and right fermion number charges (i.e., quantum numbers under the $U(1)$ R-symmetries embedded in the left/right $SU(2)$ worldsheet R-symmetries guaranteed to exist by $(4,4)$ superconformal invariance)
via
$$J_1 = F_L + F_R$$
$$J_2 = F_L - F_R$$
Counting supersymmetric solutions in space-time should be related to studying world-volume BPS states where the
right-movers are kept in a Ramond ground state.
The supersymmetric black holes constructed in supergravity have $|J_1| = |J_2| = J$, but it was already mentioned in 
\cite{BMPV} that from the effective string perspective, one can keep track of the $F_R$ quantum numbers of Ramond ground
states for the right movers.  So one should expect a more refined counting to allow independent tracking of the $F_R$ quantum number,
and hence two angular momenta.  It is this counting that we detail in this paper.  

It is important to stress that,
as noted in \cite{BMPV}, because the
right-moving Ramond ground states have $F_R$ bounded by the central charge of the effective string, this additional quantum
number can be viewed as small hair on top of the macroscopic supergravity quantum numbers.  In fact, given a count of BPS
states at some point in moduli space, only some of these will become 
microstates of a single-center black hole as one increases the string coupling.  Others may be related to hair which dresses the horizon, to multi-center solutions,
or even to solutions of distinct horizon topology.  For basic calculations extracting black hole microstate counts from BPS state counts by subtracting 
counts of the hair, see \cite{Sameer,Senhair,Murthyhair}; for a discussion of the fact that one has vanishing angular momentum for
4d black hole microstates, see \cite{zero} and references therein.  In light of these developments, it is important to stress that our conjectural BPS counting function is a count of the full spectrum of BPS states, without considerations of hair versus black hole microstates.

There is also a second development in the literature that suggests the existence of a more general counting function.  The black holes above generically have both $Q_F$ and $Q_H$, but you may count the subsector of spinning black holes only charged under one; for example, only having $Q_F$. This subsector corresponds in four dimensions to only counting the states with $Q_m^2 = 0$ (but generically carrying both $Q_e^2$ and $Q_e \cdot Q_m$ charges).  The generating function for degeneracies of such special spinning 5d black holes
(with $|J_1| = |J_2|$)
was computed by Katz, Klemm, and Vafa (KKV) in \cite{KKV} (in the 4d picture, the angular momentum becomes $Q_e \cdot Q_m$).  A refinement to account for the possibility of turning on $|J_1| \neq| J_2|$
was described recently by Katz, Klemm and Pandharipande (KKP) \cite{KKP}.  It gives rise to a three-variable generating function, 
but clearly describes black holes not carrying $Q_m^2$ (or its 5d lift).  From this perspective as well, it seems that there is a 
natural role for a more general function allowing $Q_m^2 \neq 0$, but keeping the general angular momenta allowed by the
KKP generating function.  We will construct the four-variable generating function $\Phi$ which correctly reduces to both the
DVV counting function when  one turns off the second angular momentum quantum number, and the KKP counting function
when one turns off $Q_m^2$.

 \subsection{Mathematical questions}

These counts have also been of interest mathematically in the curve-counting literature.  Depending on the particular U-duality frame one chooses, it can be more natural to interpret them in terms of Gromov-Witten invariants arising from the topological
string, Gopakumar-Vafa invariants, or Donaldson-Thomas invariants. The connections between these invariants are described
in detail in \cite{GV,INOV,Kapustin} in the physics literature, and in \cite{MNOP} in the mathematics literature.  
All of the BPS state counts above have interpretations in one or more of these frames and therefore lend themselves to mathematical interpretation. The original $1/2$-BPS state count may be interpreted as the genus $0$ Gromov-Witten theory of $K3$, as formulated in Yau-Zaslow \cite{Yau-Zaslow} and proven in \cite{KMPS}.  The partially flavored KKV count of spinning black holes corresponds to turning on a chemical potential tracking the source genus in the topological string; interpreted as the full Gromov-Witten theory of $K3$, this statement is discussed and proven in work of Pandharipande-Thomas \cite{PT}. More recently, the $1/4$-BPS state counts of $K3 \times T^2$ were interpreted as a precise mathematical conjecture in terms of Gromov-Witten invariants in \cite{OP} and discussed further in terms of Donaldson-Thomas invariants in \cite{Bryan}. Refining further corresponds to considering motivic invariants in the sense of \cite{KS}; indeed, the connection between flavoring with respect to a little group spin and considering the motivic lift of Donaldson-Thomas invariants is explained via wall-crossing in Dimofte-Gukov \cite{Gukov}.  It follows that the geometric
interpretation of the KKP counting function is as the full motivic generating function for $K3$. This leaves open the question of the full motivic curve counts for $K3 \times T^2$ (though some comments appear in the final section of 
\cite{OP}).  The four-variable function $\Phi$ we compute here constitutes a physically motivated proposal for these counts.

A perhaps surprising feature of these counting functions (sometimes explained through S-duality, which itself is poorly understood) is that they are related to automorphic functions.
It is not difficult to see that the unflavored $1/2$-BPS count is modular for the usual modular group $SL(2, \mb{Z})$; it is more interesting that the KKV and KKP counts are Jacobi and multivariable Jacobi forms, respectively. That the $1/4$-BPS state count -- which yields the Igusa cusp form --  is automorphic for the group $Sp(4, \mb{Z})$ or $O(2, 3, \mb{Z})$ is still more surprising, and many attempts have been made to explain this curious automorphy, notably including \cite{Gaiotto} and \cite{Dabholkar}. We offer no new interpretations of the automorphy of the BPS state counts in this paper, merely noting that the $1/4$-BPS state counts very often seem to be a lift of the $1/2$-BPS state counts in a precise mathematical sense; 
this holds not just for $K3\times T^2$ compactification, but for a wide variety of 4d $N=4$ supersymmetric models obtained
as so-called CHL strings (which involve quotients of $K3 \times T^2$ by a Nikulin involution on $K3$ and a shift on the torus)
\cite{Sen}.
A suitable version of our function $\Phi$ appears to be automorphic for $O(2, 4, \mb{Z})$, which again requires
physical explanation; we will not make further comment on this curious automorphy here.

There is also a connection to wall-crossing phenomena.
The refined BPS counts  jump as we move in the moduli space of our theory, undergoing wall-crossing transformations (as discussed for example in \cite{KS} or \cite{Gukov}). In fact, the wall-crossing for the dyon counting function of DVV was carefully investigated in \cite{Cheng} to establish the validity of the formula not just in one chamber of moduli space but throughout all of moduli space.  It likely will be interesting to carry out the same analysis for our refined count $\Phi$ to match the wall-crossing for the refined BPS index on the one hand with the mathematical jumping of the coefficients in our (conjecturally) automorphic function $\Phi$ as we cross walls on its domain. 

A final connection of mathematical interest is to moonshine.  Some numerical coincidences suggested to the authors of \cite{KKP} that their invariants may exhibit an analogue of Mathieu moonshine \cite{EOT}.  This story was developed in \cite{Kachru}, where the KKP invariants were re-expressed as traces in a moonshine module for Conway's largest sporadic group.
Our further refined function $\Phi$ should also naturally be expected to play a role in moonshine; we leave this connection for future work.

 The plan of the paper is as follows.  In the next section, we review the known formulae for BPS state counts before going on to state an analogous formula for our refined state-counting function $\Phi$. 
In section $3$, we define the Hodge-elliptic genus.  Section $4$ proposes a formula for the refined counting function $\Phi$ in terms of the Hodge-elliptic genus, by considering the D1-D5 system in type IIB string theory and applying an argument following the strategy of Dijkgraaf, Moore, Verlinde, and Verlinde (DMVV) \cite{DMVV}. Finally, section $5$ has a discussion of the interpretation of these results in the type IIA D0-D2-D6 frame; mathematicians will likely be most interested in this section.

\section{Statements of BPS state and black hole counts}

We recall in this section many of the counts previously computed. First, if we denote by $c_n$ the number of $1/2$-BPS states with charge squaring to $2n - 2$, we may easily compute the generating function $$f(q) = \sum c_n q^{n-1},$$ for example by working in the dual frame given by heterotic string theory compactified on $T^6$. Here, the $1/2$-BPS states are ground states on the right and hence simply given by excitations of the $24$ left-moving bosonic oscillators \cite{Harvey}. One may hence easily compute $$f(q) = \frac{1}{\eta(\tau)^{24}} = \frac{1}{\Delta(\tau)},$$ where $q = e^{2 \pi i \tau}$ and the $\eta$ function is given by $$\eta(\tau) = q^{1/24} \prod_{n=1}^{\infty} (1 - q^n).$$ 

Starting to flavor by the angular momenta, we next have the KKV count, where we write the generating function in terms of $c^r_n$, the number of $1/2$-BPS multiplets with $SU(2)_L$ representation labeled by spin $r$ and charge whose square is $2n - 2$; now, we write the generating function as $$\sum c^r_n q^{n-1} y^{[r]} = \frac{1}{\phi_{KKV}(\tau, z)},$$ where we use the notation $$j^{[\ell]} = j^{-2 \ell} + j^{-2(\ell - 1)} + \cdots + j^{2(\ell - 1)} + j^{2 \ell}.$$ We now have $\phi_{KKV}$ a Jacobi form in terms of $\tau, z$ under the relations $q = e^{2 \pi i \tau}, y = e^{2 \pi i z}$. In fact is the named Jacobi form 
$$\phi_{10,1}(\tau, z) = (y^{1/2}-y^{-1/2})^2  ~q  \prod_{n=1}^{\infty} (1 - q^n)^{20} (1 - q^n y)^2 (1 - q^n y^{-1})^2,$$
where the subscripts refer to the weight and index of the Jacobi form.

 Finally, the fully flavored count for states with $Q_H=0$, if we denote $c^{r_L, r_R}_n$ the number of representations now with spin $r_L$ under $SU(2)_L$ and $r_R$ under $SU(2)_R$, is given by $$\sum c^{r_L, r_R}_n q^{n-1} y^{[r_L]} u^{[r_R]} = \frac{1}{\phi_{KKP}(\tau, z, \nu)},$$ where we have now also introduced $u = e^{2 \pi i \nu}$ and $\phi_{KKP}$ is the multivariable Jacobi form given by 
\begin{eqnarray}
\nonumber
\phi_{KKP}(\tau, z, \nu) &=& q~ (y^{1/2}u^{1/2}-y^{-1/2}u^{-1/2})(y^{-1/2}u^{1/2}-y^{1/2}u^{-1/2}) \\
\nonumber
&&\prod_{n=1}^{\infty} (1 - q^n)^{20} (1 - q^n uy) (1 - q^n uy^{-1})
 (1 - q^n u^{-1}y) (1 - q^n u^{-1}y^{-1})
\end{eqnarray}

Moving to the $1/4$-BPS state counts, if we denote by $c_{n, \ell, m}$ the number of multiplets with charges satisfying $$\frac{1}{2} Q_e \cdot Q_e = n, Q_e \cdot Q_m = \ell, \frac{1}{2} Q_m \cdot Q_m = m,$$ we have the generating function $$\sum c_{n, \ell, m} p^{n} y^{\ell} q^{m} = \frac{1}{\Phi_{10}(\sigma, z, \tau)}.$$ We have now also introduced $p = e^{2 \pi i \sigma}$ and in fact $\Phi_{10}(\sigma, z, \tau)$ is the (unique) Siegel cusp form of degree $2$, weight $10$, and full level. We have a similar product-type formula for this generating function, given by the following multiplicative lift: 
$$\Phi_{10}(\sigma, z, \tau) = pqy \prod_{(n,m,\ell) > 0} (1 - p^n y^{\ell} q^m)^{c(nm, \ell)}~.$$
Here, $(n,m,\ell) > 0$ means $n, m \ge 0$ and $\ell$ runs over all integers, with the caveat that if $n = m = 0$, then $\ell < 0$. 
The $c(m, \ell)$ are the coefficients of the elliptic genus $$Z_{EG}(K3)(\tau, z) = \sum c(m, \ell) q^m y^{\ell},$$ with the elliptic genus being independently defined as 
$$Z_{EG}(\tau, z) = \tr_{RR} \Big( (-1)^{F_L + F_R} y^{F_L} q^{L_0 - {c\over 24}} \bar q^{\bar L_0 - {\bar c \over 24}} \Big).$$ 
$F_L, F_R$ are the left and right fermion numbers and $L_0, \bar L_0$ are the left and right moving Hamiltonians, with the trace being taken over the entire Ramond-Ramond sector of the theory. Note however that due to the $(-1)^{F_R}$ insertion, the trace localizes to the right Ramond ground states, as all other states pair up and annihilate in this index. This observation is crucial to the extension we define next. Note furthermore that just as the $1/4$-BPS state count was a multiplicative lift of the elliptic genus in the sense that the coefficients of the elliptic genus appear as exponents in a product formula for the generating function, the $1/2$-BPS state counts are also multiplicative lifts of simpler expressions, namely the (symmetrized) Euler characteristic, $\chi_y$ genus, and Hodge polynomial of $K3$ respectively.  Explicitly, these are given by \begin{eqnarray*} \chi(K3) &=& 24 \\ \chi_y(K3) &=& 2y^{-1} + 20 + 2y \\ \chi_{Hodge}(K3) &=& u^{-1}y^{-1} + u^{-1}y + 20 + uy^{-1} + uy. \end{eqnarray*} We will return to this observation in section $4$ when we revisit the DMVV argument; for now, we simply comment that we will express the refined count of $1/4$-BPS states in terms of a multiplicative lift of an interpolation between the Hodge polynomial and the elliptic genus, which we term the Hodge-elliptic genus.

\section{The Hodge-elliptic genus}

The definition is simple: $$Z_{HEG}(\tau, z, \nu) = \tr \Big((-1)^{F_L + F_R} y^{F_L} u^{F_R} q^{L_0 - {c \over 24}} \Big),$$ where now the trace is crucially taken over only those states whose right-moving part is a Ramond ground state. This definition makes it unclear that we have an index, i.e. that this quantity is invariant under deformation, and indeed, this quantity should jump in a prescribed semicontinuous fashion as we vary in moduli space for a general Calabi-Yau. In fact, it is manifest from the definition that wherever the (left) chiral algebra enhances as one moves in CFT moduli space, the
formula will jump -- at those points, extra states exist with no right-moving excitation.
An index related to ours -- roughly, $\left({\partial^2 \over \partial \nu^2} Z_{HEG}\right) \vert_{\nu = 0}$ -- was studied previously in \cite{MMS}.

%The extra supersymmetry in our theory, however, protects against such behavior (SAY SOME SHIT ABOUT HYPERKAHLER). 

We can also propose a purely mathematical definition.  
Recall that the elliptic genus \cite{Warner,EGrefs,Yamada} may be reformulated as a holomorphic Euler characteristic as follows: \begin{multline*} Z_{EG}(\tau, z. \nu) = i^{r - D} q^{(r - D) / 12} y^{-r / 2} \chi \Big( X, \bigotimes_{n=1}^{\infty} \Lambda_{-yq^{n-1}} \mc{E} \otimes \bigotimes_{n=1}^{\infty} \Lambda_{-y^{-1}q^n} \mc{E}^{\vee} \\ \otimes \bigotimes_{n=1}^{\infty} \Sym_{q^n} TX \otimes \bigotimes_{n=1}^{\infty} \Sym_{q^n} T^*X \Big). \end{multline*} Here we give the formula relevant for a general (0,2) supersymmetric CFT, whose specifying data
includes a holomorphic bundle $\mc{E}$ on $X$, where we use the notation $$\Lambda_t \mc{E} = \bigoplus_{s = 0}^{\infty} t^s \Lambda^s \mc{E}$$ and $$\Sym_t \mc{E} = \bigoplus_{s = 0}^{\infty} t^s \Sym^s \mc{E}$$ for formal power series of bundles. Here $r$ denotes the (complex) dimension of $X$ and $D$ the rank of the bundle $\mc{E}$, while $\chi(X, -)$ denotes the holomorphic Euler characteristic for (formal power series of) holomorphic bundles $$\chi(X, \mc{E}) = \sum_{j = 0}^r (-1)^j \dim H^j(X, \mc{E}).$$

Following the same reasoning, we may upgrade the above formula to one for the Hodge-elliptic genus as follows: \begin{multline*} Z_{HEG}(\tau, z, \nu) = i^{r - D} q^{(r - D) / 12} y^{-r / 2} \sum_{j=0}^r (-u)^j \dim H^j \Big( X, \bigotimes_{n=1}^{\infty} \Lambda_{-yq^{n-1}} \mc{E} \otimes \bigotimes_{n=1}^{\infty} \Lambda_{-y^{-1}q^n} \mc{E}^{\vee} \\ \otimes \bigotimes_{n=1}^{\infty} \Sym_{q^n} TX \otimes \bigotimes_{n=1}^{\infty} \Sym_{q^n} T^*X \Big). \end{multline*} Indeed, mathematicians may take this as a definition of the Hodge-elliptic genus in some level of generality. 

As an example, it is easy to compute the Hodge-elliptic genus of an abelian variety starting from this expression: adopting the convention as usual that if the holomorphic bundle $\mc{E}$ is not specified, we take it to be the tangent bundle $TX$, the triviality of the tangent bundle of an abelian variety implies we are merely taking power series in the trivial bundle. In particular, the ranks of the cohomology of the trivial bundle do not depend on the complex moduli of our abelian variety by Hodge theory as $$\dim H^j(X, \mc{O}) = h^{0, j}(X)$$ does not vary in complex moduli. As such, we may assume our abelian variety is a product of elliptic curves; just as with the elliptic genus, the Hodge-elliptic genus is multiplicative, allowing us to reduce to the case of an elliptic curve. Finally, we have for an elliptic curve $E$ that \begin{eqnarray*} Z_{HEG}(E)(\tau, z, \nu) &=& y^{-1/2} \Big( \sum_{j=0}^1 (-u)^j \dim H^j(E, \mc{O}_E) \Big) \times \\ && \prod_{n=1}^{\infty} (1 - yq^{n-1}) \prod_{n=1}^{\infty}(1 - y^{-1}q^n) \Big(\prod_{n=1}^{\infty} (1 + q^n + q^{2n} + \cdots)\Big)^{2} \\ &=& y^{-1/2}u^{-1/2} (1 - u) \prod_{n=1}^{\infty} \frac{(1 - yq^{n-1})(1 - y^{-1}q^n)}{(1 - q^n)^2} \\ &=& 4 \frac{\theta_1(\tau, z)}{\theta_1^*(\tau, 0)} u_-,\end{eqnarray*} where we will use the notation $$u_{\pm} = \frac{u^{-1/2} \pm u^{1/2}}{2}$$ and the Jacobi theta function $$\theta_1(\tau, z) = - q^{1/8} y^{-1/2} \prod_{n=1}^{\infty} (1 - yq^{n-1}) (1 - q^n)(1 - y^{-1}q^n),$$ where we denote $$\theta_1^*(\tau, 0) = -2q^{1/8} \prod_{n=1}^{\infty} (1 - q^n)^3.$$ We hence easily have by the above reasoning that $$Z_{HEG}(T^{2r}) = \Big( 4 \frac{\theta_1(\tau, z)}{\theta_1^*(\tau, 0)} u_- \Big)^r,$$ where we just denote the dependence on the underlying manifold $T^{2r}$ as we have already argued the complex structure of the $r$-dimensional abelian variety is irrelevant. 

Returning to the case of $K3$, we may also argue from the expression in terms of cohomology ranks that the Hodge-elliptic genus is invariant under deformation of complex structure. Recall from Bochner's formula that global sections of powers of the tangent bundle of a Kahler-Einstein manifold are necessarily parallel tensor fields, i.e. invariant under parallel transport and hence determined by the invariance of the holonomy representation on the fibre at the identity; see for example \cite{Kobayashi} for an overview. As such, the rank of the global sections in the formula above are completely determined by linear algebra and hence obviously deformation-invariant; similarly, by Serre duality, the same argument applies to the ranks of the second cohomology groups and as the holomorphic Euler characteristics are necessarily invariant in flat families, the same must be true for the first cohomology ranks, showing that the Hodge-elliptic genus is in fact deformation-invariant for any Kahler-Einstein surface as expressed above.

It is important to note that the purely mathematical definition above follows from the trace definition in a CFT only for
CFTs which describe large radius sigma models with target $X$.  Away from large radius, as noted previously, the chiral algebra
of the sigma model can be enhanced (on loci of proper codimension) by additional chiral currents.  (These will always be of higher spin for compact Calabi-Yau targets, as such spaces admit no continuous isometries).  This will lead to jumps in the trace definition of the genus; the mathematical definition only captures
the strict large-volume limit.  The jumps cancel in the limit that one takes $u \to 1$ because the chiral currents come paired
with superpartners that then cancel in the index; this recovers the constancy of the elliptic genus.  Away from points with
enhanced chiral algebra -- i.e., at generic points in the moduli space of superconformal theories -- the Hodge-elliptic genus
will be constant on the moduli space.

We close by giving the result of a sample calculation.  We choose a particularly simple point in moduli space  --  a Kummer orbifold point where $K3$ is considered as a resolution of $T^4$ by the inversion $\mb{Z}/2$-action (and where the $T^4$ is a product of two square $T^2$s with unit volume). Just as the elliptic genus of $K3$ may be expressed in terms of theta functions as 
\cite{EOTY}
$$Z_{EG}(K3)(\tau, z) = 8 \Big( \Big( \frac{\theta_2(\tau, z)}{\theta_2(\tau, 0)} \Big)^2 + \Big( \frac{\theta_3(\tau, z)}{\theta_3(\tau, 0)} \Big)^2 + \Big( \frac{\theta_4(\tau, z)}{\theta_4(\tau, 0)} \Big)^2 \Big),$$ the Hodge-elliptic genus is now given as a sum over all four sectors of $\mb{Z}/2$-periodicity conditions we may put on the torus (morphisms from its fundamental group to $\mb{Z}/2$), as now the right-moving fermion zero-mode no longer causes the first contribution to cancel. In fact, we calculate $$Z_{HEG}(K3)(\tau, z, \nu) = 8 \Big( \Big( \frac{\theta_1(\tau, z)}{\theta_1^*(\tau, 0)} u_- \Big)^2 + \Big( \frac{\theta_2(\tau, z)}{\theta_2(\tau, 0)} u_+ \Big)^2 + \Big( \frac{\theta_3(\tau, z)}{\theta_3(\tau, 0)} \Big)^2 + \Big( \frac{\theta_4(\tau, z)}{\theta_4(\tau, 0)} \Big)^2 \Big).$$ 
This is the value of the Hodge-elliptic genus for a particularly symmetric orbifold K3.  A conjecture
for the form of the genus for ${\it generic}$ values of the K3 moduli, where there is no enhanced
worldsheet chiral algebra, has been put forward by Katrin Wendland \cite{Wendland}.
We return to the importance of $Z_{HEG}(K3)$ in the next section.

\section{The D1-D5 system and symmetric powers}

We now explain how to obtain the refined count of spinning BPS states from a lift of the Hodge-elliptic genus.  First, we revisit the DMVV argument calculating the free energy of a gas of D1 branes dissolved inside a D5 wrapping $K3 \times S^1$ in a IIB compactification, where the D1s wrap the auxiliary $S^1$ and are localized to points in the $K3$. As usual, ignoring the trivial dynamics of the six-dimensional $U(1)$ gauge-theory from the worldvolume of the D5, the dynamics of this system with $N$ D1s is given by a nonlinear sigma-model mapping to the orbifold symmetric power $(K3)^n$ modulo the symmetric group $S_n$-action by permutation of the factors. We will denote this orbifold by $\Sym^n K3$ but mathematicians should be aware we mean the orbifold conformal field theory as opposed to the sigma-model of maps to this singular space itself; equivalently, by deforming in the Kahler moduli space to resolve singularities, we may consider the sigma-model to the crepant resolution $\Hilb^n K3$. As such, mathematicians may mentally substitute $\Hilb^n K3$ below for all mentions of $\Sym^n K3$. In any case, DMVV are able to argue that the unrefined BPS state count in five dimensions is given by $$\sum_{n=0}^{\infty} p^n Z_{EG}(\Sym^n K3),$$ which they then compute via elegant arguments entailing how the string states in the orbifold CFTs arrange themselves into long strings in the original sigma-model, allowing for immediate multiplicative lift formulas that apply not just to the elliptic genus but also to many other indices of manifolds. Indeed, by the same arguments, we have all of the following identities: \begin{eqnarray*} \sum_{n=0}^{\infty} p^n \chi(\Sym^n K3) &=& \prod_{k=1}^{\infty} \frac{1}{(1 - p^k)^{24}} = \frac{p}{\Delta(\sigma)} \\ \sum_{n=0}^{\infty} p^n \chi_{-y}(\Sym^n K3) &=& \prod_{k=1}^{\infty} \frac{1}{(1 - yp^k)^2(1 - p^k)^{20}(1 - y^{-1}p^k)^2} = \frac{p(-y + 2 - y^{-1})}{\phi_{10, 1}(\sigma, z)} \\ \sum_{n=0}^{\infty} p^n Z_{EG}(\Sym^n K3) &=& \prod_{r > 0, s \ge 0, t} \frac{1}{(1 - q^s y^t p^r)^{c(rs, t)}} = \frac{p \phi_{10, 1}(\tau, z)}{\Phi_{10}(\sigma, z, \tau)}, \end{eqnarray*} where we define the coefficients $c(n, \ell)$ via the expansion $$Z_{EG}(K3) = \sum_{n, \ell} c(n, \ell) q^n y^{\ell}.$$ 

The last formula in particular makes clear the nature of this multiplicative lift: we use the coefficients of the index on the original manifold $K3$ as exponents in the product representation for the generating function of the index over all symmetric powers; indeed, the above two formulas are special cases but also correspond to the simple observations $\chi(K3) = 24$ and $\chi_{-y}(K3) = 2y^{-1} + 20 + 2y$, making clear where the exponents come from in the first two formulae. The reason that the automorphic representations on the right-hand side of the above product formulas require some correction factors in the numerator is due to the ranges of the parameters in the product of $r > 0$ and $s \ge 0$; those correction factors disappear if we were to instead have products over $r, s \ge 0$, as is the case if we count the four-dimensional BPS states rather than the five-dimensional BPS states. Indeed, those states with $r = 0$ are exactly those introduced (or removed) under the 4d-5d lift as the kinematic factors for the center-of-mass of the $D5$ brane in the five-dimensional frame.

The multiplicative lift property is made clear from DMVV's analysis of the Hilbert space of the symmetric power CFTs, from which all these formulas follow by simply taking traces. In fact, the arguments work equally well if we only consider the subspace of the Hilbert space given by Ramond ground states on the right, so that we may also take the traces that define the symmetrized Hodge polynomial or Hodge-elliptic genus of $K3$ and derive the following formulas: \begin{eqnarray*} \sum_{n=0}^{\infty} p^n \chi_{Hodge}(\Sym^n K3) &=& \prod_{k=1}^{\infty} \frac{1}{(1 - uyp^k)(1 - u^{-1}yp^k)(1 - p^k)^{20}(1 - uy^{-1}p^k)(1 - u^{-1}y^{-1}p^k)} \\ &=& \frac{p(u-y-y^{-1}+u^{-1})}{\phi_{KKP}(\sigma, z, \nu)} \\ \sum_{n=0}^{\infty} p^n Z_{HEG}(\Sym^n K3) &=& \prod_{r > 0, s \ge 0, t, v} \frac{1}{(1 - q^s y^t p^r u^v)^{c(rs, t, v)}} = \frac{p \phi_{KKP}(\tau, z, \nu)}{\Phi(\sigma, z, \tau, \nu)},\end{eqnarray*} where now we have the coefficients of the Hodge-elliptic genus \begin{eqnarray*} Z_{HEG}(K3)(\tau, z, \nu) &=& 8 \Big( \Big( \frac{\theta_1(\tau, z)}{\theta_1^*(\tau, 0)} u_- \Big)^2 + \Big( \frac{\theta_2(\tau, z)}{\theta_2(\tau, 0)} u_+ \Big)^2 + \Big( \frac{\theta_3(\tau, z)}{\theta_3(\tau, 0)} \Big)^2 + \Big( \frac{\theta_4(\tau, z)}{\theta_4(\tau, 0)} \Big)^2 \Big) \\ &=& \sum_{n, \ell, m} c(n, \ell, m) q^n y^{\ell} u^m.\end{eqnarray*} 
Here, we have implicitly defined our four-variable counting function
$$\Phi(\sigma, z, \tau, \nu) = pqy \prod (1 - q^s y^t p^r u^v)^{c(rs, t, v)}$$ 
as given by a multiplicative lift in terms of the Fourier coefficients $c(n,\ell,m)$ of $Z_{HEG}(K3)$. As in the DMVV argument that gave rise to the definition of $\Phi_{10}$, the index set in the product representation for $\Phi$ requires some care. The correct index set over which we take the product is over all $r, s \ge 0$ and all $t, v$ except that if $r = s = 0$, we restrict to $t < 0$.
We see that $\Phi$ is simply related to $\Phi_{10}$ when $u=1$ and $\phi_{KKP}$ when 
$q \to 0$.

To relate these symmetric power sigma-models back to the D-brane BPS state counts, we have only to note that the extra $U(1)$ charge we wish to turn on in the BPS state count is precisely the same as that which we turn on in the Hodge-elliptic genus. Then the five-dimensional refined BPS state count is precisely given by the formula above for the multiplicative lift of the Hodge-elliptic genus.
For completeness, we also note the four-dimensional count of BPS states, as differing by the $r = 0$ modes: denoting $c_{n, \ell, m, r}$ the number of four-dimensional $1/4$-BPS multiplets with charges labelled as before and spin $r$, we have $$\sum c_{n, \ell, m, r} p^n y^{\ell} q^m u^{[r]} = \frac{1}{\Phi(\sigma, z, \tau, \nu)}.$$

Technically speaking, the above product representation for $\Phi$ only converges in one chamber of moduli space and must be analytically continued elsewhere as a meromorphic function.  The BPS state counts as extracted from a contour integral procedure will hence jump as the contours cross over poles and a wall-crossing analysis will ensue.  Close to the point of definition, however, the above formula holds on the nose and gives the precise refined BPS counting function.

Here, we see explicitly that in the 5d state count, the new variable parametrizes hair which is small compared to the charges visible in supergravity.
The factor of $\phi_{KKP}$ in the numerator causes the powers of $u$ to be strictly bounded in terms of the powers of $p$: in no term that occurs in the expansion of the five-dimensional count does the exponent of $u$ exceed that of $p$, irrespective of the exponents of $q$ or $y$. This fact is clear from the original definition of the Hodge-elliptic genus but also is necessary for this count of spinning black holes to be physically meaningful. Indeed, in order to refine this count by extra hair such as this new $U(1)$ charge, standard supergravity  literature informs us this hair must be microscopic, i.e. not visible at the semiclassical level; the BPS black holes in supergravity have
$|J_1| = |J_2|$.  The 4d counting formula, where the automorphic correction factor in the numerator of the 5d formula vanishes,
doesn't limit the powers of $u$ in the same way; this should likely be interpreted as counting of additional multicenter or small black hole states with large angular momentum in the 4d picture.

\section{The IIA frame}

Adding a circle and T-dualizing the D1-D5 system in type IIB, or starting from M-theory and compactifying along the M-theory circle, allows us to consider this system as a D0-D2-D6 brane configuration in IIA string theory.  The corresponding state counts have well-established mathematical interpretations as (weighted) Euler characteristics of the relevant moduli spaces. Typically we could also interpret these invariants in terms of the topological string, but unfortunately the refinement we conduct in this paper does not yet have a mathematical interpretation in the topological string: motivic Gromov-Witten invariants are yet to be defined. As such, we phrase the mathematically formulated versions of our conjectures in terms of motivic Donaldson-Thomas invariants to retain at least some degree of mathematical precision. We defer full precision and some checks to later work.

We first review the ordinary Donaldson-Thomas invariants of $K3 \times T^2$ as discussed in \cite{Bryan,Oberdieck}. Here, for $X$ the Calabi-Yau threefold given as a product of some $K3$ surface $S$ with an elliptic curve $E$, we first consider the Hilbert scheme of subschemes of $X$ with fixed numerical invariants as follows: $$\Hilb^{\beta, n}(X) = \{Z \subset X \Big| [Z] = \beta, \chi(\mc{O}_Z) = n\},$$ where $\beta \in H_2(X ; \mb{Z})$ is a curve class in $S$, $Z$ is a subscheme of $X$, and $[Z]$ represents its underlying class in homology. Note that $X$ has a natural $E$-action by translation; this action is inherited by $\Hilb^{\beta, n}(X)$ and so taking a (weighted) Euler characteristic directly would vanish, corresponding to the extra supersymmetry physically. Eliminating the superfluous fermionic zero-modes corresponds mathematically to instead taking the weighted Euler characteristic of the quotient of this Hilbert scheme by the $E$-action, and we have the reduced Donaldson-Thomas invariants $$\mathrm{DT}_{\beta, n}(X) = \int_{\Hilb^{\beta, n}(X)/E} \nu d\chi,$$ where $\nu$ is the usual Behrend function and we integrate with respect to Euler characteristic.   We now conjecture that we may define $\mathrm{DT}_{h, d, n}(X)$ as $\mathrm{DT}_{\beta + dE, n}(X)$ where $\beta$ is a (primitive) class satisfying $\beta^2 = 2h - 2$; Bryan \cite{Bryan} argues for this invariance using deformation-invariance, sketching a proof using the machinery of shifted symplectic structures. Assembling these invariants into a generating function as $$\mathrm{DT}(X) = \sum_{h, d \ge 0, n} \mathrm{DT}_{h, d, n}(X) q^{h-1} p^{d - 1} (-y)^n,$$ we in fact have the mathematical conjecture $$\mathrm{DT}(X) = -\frac{1}{\Phi_{10}(\sigma, z, \tau)}.$$

In these terms, it is clear how to extend to a motivic version, modulo some natural conjectures. First, we note following \cite{Pantev,Bussi} that the Hilbert scheme used to define Donaldson-Thomas invariants has a motivic incarnation $$DT^{mot}_{\beta, n}(X) \in K^{\hat{\mu}}(\mathrm{Var})[\mb{L}^{-1}],$$ the Grothendieck ring of varieties with action by roots of unity as extended by inverting the Tate class. Unfortunately, a difficulty arises here when choosing orientation data, roughly corresponding to the choice of spin structure on the relevant moduli spaces of sheaves. In previous work along these lines such as in \cite{KKP}, all relevant moduli spaces are conjecturally simply-connected and so there is a single choice for orientation datum. Here, we seem to have four choices, and it remains to explore their compatibility as we vary in moduli space. We make no further reference to pinning down this detail here, merely conjecturing that there exists some consistent choice such that the following conjectures hold. 

We now further conjecture a canonical quotient by the $E$-action by translation, i.e. we conjecture a natural reduced motivic Donaldson-Thomas invariant related to the above via $$ DT^{mot}_{\beta, n}(X) = \mathrm{DT}^{mot}_{\beta, n}(X) \cdot [E]$$ as an equation of classes in $K^{\hat{\mu}}(\mathrm{Var})[\mb{L}^{-1}]$. Taking the motivic measure given by the Euler characteristic should now return the reduced Donaldson-Thomas invariants; if we instead take the (symmetrized) Poincar\'{e} polynomial, which we denote by $$P_{\beta, n}(X) = P(\mathrm{DT}^{mot}_{\beta, n}(X)) \in \mb{Q}[u, u^{-1}],$$ we now further conjecture that this Laurent polynomial only depends on $h, d,$ and $n$, where as before we take a curve class $\beta + d E$ for $\beta$ a curve class in $K3$ satisfying $\beta^2 = 2h - 2$. Assuming these conjectures, we may assemble the generating function $$\mathrm{DT}^{mot}(X) = \sum_{h, d \ge 0, n} P_{h, d, n}(X) q^{h-1} p^{d - 1} (-y)^n,$$ which we conjecture should be given by $$\mathrm{DT}^{mot}(X) = -\frac{1}{\Phi(\sigma, z, \tau, \nu)}$$ for $\Phi$ the multiplicative lift of $Z_{HEG}(K3)$ as above.

In \S3, we saw that the natural field theory definition of the Hodge-elliptic genus has `jumps' at points in moduli space with
enhanced chiral algebra (while admitting a `generic' answer that is constant away from such points).   This implies a similar
jumping phenomenon for the motivic Donaldson-Thomas invariants, which has been seen explicitly in computations on 
non-commutative Calabi-Yau threefolds \cite{Jump}. In particular, varying in the complex or K\"{a}hler moduli space should correspond to varying the algebraic structure or the stability structure on the derived category, which may \textit{a priori} lead to different motivic Donaldson-Thomas counts. Our conjecture implies that the jumping behavior of these refined curve counts should match precisely to where the conformal field theory enjoys extra currents as we vary in the K3 $\sigma$-model moduli space. Following~\cite{Wendland}, the computation $Z_{HEG}$ for a generic K3 $\sigma$-model is now known following certain natural assumptions, and so lifting this function should give the motivic Donaldson-Thomas invariants for generic complex and K\"{a}hler parameters. The four-variable function $\Phi$ we give earlier is for a generic K3 surface of Kummer type, and so we expect the corresponding lift to give the motivic Donaldson-Thomas invariants (with the correct choice of orientation datum) for precisely these surfaces.

\bigskip
\centerline{\bf{Acknowledgements}}
\medskip
We would like to thank G. Oberdieck, N. Paquette, C. Vafa, R. Vakil, R. Volpato, and K. Wendland
for helpful conversations, and K. Wendland of informing us about her conjecture for the generic value
of $Z_{HEG}(K3)$.  We are especially grateful to S. Murthy and A. Sen for discussions of
BPS state counts and black hole hair, and to D. Maulik and G. Oberdieck for helpful comments on the role of orientation data in defining motivic Donaldson-Thomas invariants. S.K. was supported in part by the National Science Foundation under grant
PHY-1316699.

\end{document}